\documentclass[a4paper,twocolumn,floatfix]{revtex4}
\usepackage[utf8]{inputenc}
\usepackage{hyperref}
\usepackage{color}
\usepackage[normalem]{ulem}
\usepackage{graphicx}
\usepackage{amsmath}
\newcommand{\eps}{\varepsilon}
\newcommand{\kF}{k_{\text{F}}}

\begin{document}
\title{Effect of the equation of state for dilute neutron matter on the composition of the inner crust of neutron stars}

\author{Sunny Kumar Gupta}
\email{sunny20@iiserb.ac.in}
\affiliation{Indian Institute of Science Education and Research Bhopal, Bhopal 462066, India}

\author{Michael Urban}
\email{michael.urban@ijclab.in2p3.fr}
\affiliation{Universit\'e Paris-Saclay, CNRS-IN2P3, IJCLab, 91405 Orsay, France}

%\date{December 2023}
\begin{abstract}
The composition of the inner crust of neutron stars is usually studied using phenomenological interactions such as Skyrme energy-density functionals. But most of these functionals do not agree well with ab-initio calculations of very dilute neutron matter. 
In this work, we study the inner crust of neutron stars in the model of phase coexistence of dense neutron-rich nuclear clusters and dilute neutron gas, and we investigate how employing a realistic microscopic equation of state to the neutron gas alters the composition. Our results indicate that with a functional that reproduces the correct equation of state of neutron matter at moderate densities, one can obtain a good description of the crust even if the functional does not have the correct behavior at extremely low density.
\end{abstract}
\maketitle

%%%%%%%%%%%%%%%%%%%%%%%%%%%%%%%%%%%%%%%%%%%%%%%%%%%%%%%%%%%%%%%%%%%%%%%%%%%%%%%%%%%%%%%%%%%%%%%%%%%%%%%%%%%%%%%%%%%%%%%%%%%%
\section{Introduction}
%%%%%%%%%%%%%%%%%%%%%%%%%%%%%%%%%%%%%%%%%%%%%%%%%%%%%%%%%%%%%%%%%%%%%%%%%%%%%%%%%%%%%%%%%%%%%%%%%%%%%%%%%%%%%%%%%%%%%%%%%%%%
Neutron stars are born in core-collapse supernovae and are among the most compact objects found in
the universe. The presence of dense matter and its extreme physical properties make neutron stars a cosmic laboratory to
study fundamental laws under conditions inaccessible in terrestrial
experiments. The recent detection of gravitational wave signals of binary neutron star mergers \cite{AbbottGW2017}
sparked great interest among the new generation of physicists to study the physical properties and composition of neutron
stars. Neutron stars consist of four layers, namely outer crust, inner crust, outer core, and inner core \cite{Chamel2008}.
The inner crust is composed of bound nuclear clusters dipped in less dense superfluid neutrons
and degenerate electron gas. Determination of its properties and composition holds importance in understanding exotic
phases of matter (the so-called ``pasta-phases''), and in explaining pulsar glitches which are said to be the
consequence of the effect of the pinning of vortices of the superfluid neutrons to the lattice of clusters \cite{Anderson1975}.

\begin{figure}
    \includegraphics[width = \linewidth]{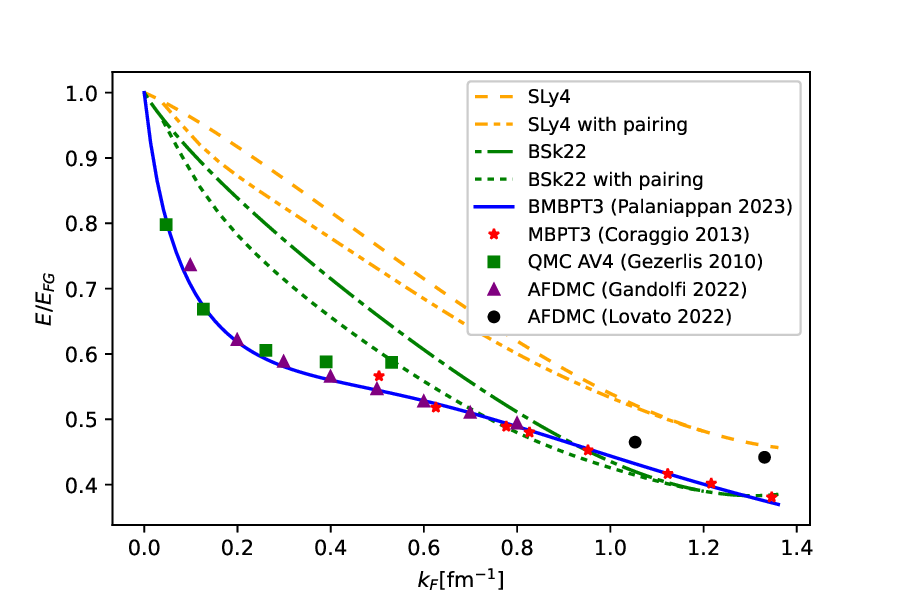}
    %\new{[Explain all lines and add symbols for QMC]}
    \caption{Comparison of low-density neutron-matter EoS obtained from phenomenological Skyrme EDF (SLy4 (orange dashes) \cite{Chabanat1997}  and BSk22 (green dash-dot) \cite{Chamel2009,Chamel2013})  with various EoS obtained from realistic $nn$ interactions: third-order BMBPT (blue solid line) \cite{Palaniappan2023}, MBPT (red stars) \cite{Coraggio2013}, and various QMC calculations \cite{Gezerlis2010} (green squares), \cite{Gandolfi2022} (purple triangle) and \cite{Lovato2022} (black circles). The vertical axis represents the ground-state energy in units of the energy of the free Fermi gas ($E_{\text{FG}}$), as a function of Fermi momentum ($\kF$).}
    \label{fig:comparison}
\end{figure}

The inhomogeneous phase of the inner crust depends on the equation of state (EoS) of low-density neutron-rich matter.
In a first approximation, it can be regarded as a phase coexistence of liquid drops (nuclear clusters) and a gas
(dilute neutron gas) under the constraints of charge neutrality and $\beta$-equilibrium \cite{Avancini2008,Martin2015,Grams2017}. However, the
results depend upon the choice of phenomenological interactions like Skyrme energy-density functionals (EDF) such as SLy4 \cite{Chabanat1997}, BSk22 \cite{Chamel2009,Chamel2013}, or relativistic mean-field 
models \cite{Pais2014}. It has been known for long that if Skyrme EDFs are fitted only to bound nuclei, they fail to correctly describe neutron matter \cite{Negele1973,Pethick1995}. Therefore, more modern Skyrme forces such as SLy4 \cite{Chabanat1997} or BSk22 \cite{Chamel2013} always include neutron-matter constraints into their fitting protocol. In spite of this, they fail to correctly reproduce the EoS of the neutron gas at very low densities, as noticed in \cite{Yang2016}. As an example, we show in Fig.~\ref{fig:comparison} the neutron-matter EoS as predicted by
the SLy4 (orange dashes) and BSk22 (green dash-dot) Skyrme EDFs compared to theories developed for infinite neutron matter using realistic neutron-neutron ($nn$) interactions as starting point (the so-called ab-initio theories), namely, various Quantum-Monte-Carlo (QMC) calculations \cite{Gezerlis2010,Lovato2022,Gandolfi2022} and recent calculations within Many-Body Perturbation Theory (MBPT, red stars) \cite{Coraggio2013} and Bogoliubov Many-Body Perturbation Theory (BMBPT, blue solid line) \cite{Palaniappan2023} using renormalization-group evolved interactions.
%%%%%%%%%%%%%%%%%%%%%%%%%%%%%%%%%%%%%%%%%%%%%%%%%%%%%%%%%%%%%%%%%%%%%%%%%%%%%%%%%%%%%%%%%%%%%%%%%%%%%%%%%%%%%%%%%%%%%%%%%%%%
%%%%%%%%%%%%%%%%%%%%%%%%%%%%%%%%%%%%%%%%%%%%%%%%%%%%%%%%%%%%%%%%%%%%%%%%%%%%%%%%%%%%%%%%%%%%%%%%%%%%%%%%%%%%%%%%%%%%%%%%%%%%
We see that there is a huge discrepancy between the Skyrme and the ab-initio results in the low-density regime. In the  case of BSk22, the discrepancy is essentially limited to the density range below $\sim 0.02$ fm$^{-3}$. One may expect that including $nn$ pairing on top of the Skyrme mean field would make the agreement better, since the pairing energy lowers the neutron-matter energy in this density range \cite{Zhang2019}. However, as shown in Fig.~\ref{fig:comparison}, adding the pairing energy (see Appendix \ref{app:pairing} for details) to SLy4 (orange dash-dot-dot) or BSk22 (green dots), is not sufficient to bring the phenomenological EoS into agreement with ab-initio results. In the rest of this paper, we will neglect the pairing energy.

Many previous studies of the inner crust relied entirely on the phenomenological Skyrme EDF. These studies include, e.g., Hartree-Fock \cite{Negele1973} and Hartree-Fock-Bogoliubov \cite{Shelley2020} calculations, and simpler extended Thomas-Fermi calculations \cite{Martin2015} to which shell corrections can be added perturbatively \cite{Pearson2012}. Looking at Fig. \ref{fig:comparison}, one may wonder how reliable these descriptions can be if they use functionals that do not have the right low-density limit. To
address this question, we employ the simple phase coexistence model \cite{Avancini2008,Martin2015,Grams2017}
neglecting surface energy, Coulomb energy, and shell effects, and minimizing the thermodynamic potential using the phenomenological EDF for the drops and the realistic neutron-matter EoS for the gas. 

%%%%%%%%%%%%%%%%%%%%%%%%%%%%%%%%%%%%%%%%%%%%%%%%%%%%%%%%%%%%%%%%%%%%%%%%%%%%%%%%%%%%%%%%%%%%%%%%%%%%%%%%%%%%%%%%%%%%%%%%%%%%
\section{Formalism}
%%%%%%%%%%%%%%%%%%%%%%%%%%%%%%%%%%%%%%%%%%%%%%%%%%%%%%%%%%%%%%%%%%%%%%%%%%%%%%%%%%%%%%%%%%%%%%%%%%%%%%%%%%%%%%%%%%%%%%%%%%%%
The inner crust contains neutron gas, bound nuclei, and electron gas. Following \cite{Martin2015}, it can be modelled as phase coexistence of liquid drops (bound nuclei) with volume $V^{(l)}$, neutron gas with volume $V^{(g)}$ and uniformly distributed electron gas to satisfy charge neutrality in the total volume,
$V = V^{(l)} + V^{(g)}$,
where the liquid volume fraction is
    $u = V^{(l)}/{V}$.
Charge neutrality implies
     $n_{e} = u n_{p}^{(l)}$,
where $n_{q}$ is the number density of particle species $q = n,p,e$.
The total energy of the system is given by $E = \eps V$ where $\eps$ is the total (average) energy density. In this simple model, the
total energy density as a function of $ u, n_{p}^{(l)}, n_{n}^{(l)}$ and $n_{n}^{(g)}$ is
\begin{equation}
    \eps = \eps_{e}(un_{p}^{(l)}) + (1-u)\eps_{N}(n_{n}^{(g)},0) + u \eps_{N}(n_{n}^{(l)},n_{p}^{(l)})\,,
    \label{energydensity}
\end{equation}
where $\eps_{e}(n_{e}) = \tfrac{3}{4}n_{e}\hbar k_{F,e} $ is the energy density of the electron gas (neglecting the electron mass) and
$\eps_{N}(n_n,n_p)$ denotes the energy density of nuclear matter with neutron and proton densities $n_n$ and $n_p$ respectively.
The total baryon density is again a function of $ u, n_{p}^{(l)}, n_{n}^{(l)}$ and $n_{n}^{(g)}$ given by
\begin{equation}
    n_{B} = un_{p}^{(l)}+un_{n}^{(l)}+(1-u)n_{n}^{(g)}\,.
    \label{eq:nB}
\end{equation}

To stabilise the system, it should have minimum energy under the constraint of constant baryon density, therefore minimizing the function
$\eps - \lambda n_{B}$,
where $\lambda$ is a Lagrange multiplier which eventually will turn out to be the baryon chemical potential $\mu_{B}$. 
By doing so we get three equations which describe mechanical equilibrium, chemical equilibrium and $\beta$-equilibrium,
\begin{gather}
    P^{(l)} = P^{(g)}\,,\label{pressure}\\
    \mu_{n}^{(l)} = \mu_{n}^{(g)}\,,\label{fixmub}\\
    \mu_{n}^{(l)} = \mu_{e} + \mu_{p}^{(l)}\,.\label{beta}
\end{gather}
In phase $i=l,g$, the pressure $P^{(i)}$ is given by,
\begin{equation}
   P^{(i)} = \mu_{n}^{(i)} n_{n}^{(i)} + \mu_{p}^{(i)} n_{p}^{(i)} - \eps(n_{n}^{(i)},n_{p}^{(i)}) \, ,
\end{equation}
and the chemical potential by $\mu_{q}^{(i)} = \partial \eps/\partial n_{q}^{(i)}$.
For given $n_n^{(g)}$ (or equivalently, for given $\mu_n^{(g)}$), we can solve numerically Eqs. (\ref{pressure}) and (\ref{fixmub}) for the variables $n_{p}^{(l)}$ and $n_{n}^{(l)}$. Then, the volume fraction can be found from
\begin{equation}
    u = {\biggr[ \frac{\mu_{n}^{(l)} - \mu_{p}^{(l)}}{\hbar c} \biggr]}^3 \frac{1}{3 \pi^2 n_{p}^{(l)}} \,,\label{eliminating_u}
\end{equation}
which is a simplified expression of Eq.(\ref{beta}), and finally the total baryon density is given by Eq. (\ref{eq:nB}).

In principle, the pressures and chemical potentials in the two phases follow from one and the same EoS. For instance, we will make as in Ref.
\cite{Martin2015} calculations using pressures $P^{(i)}$ and chemical potentials $\mu_q^{(i)}$ obtained with the SLy4 and the BSk22 parametrizations
of the Skyrme EDF. However, knowing that the Skyrme EDF fails to describe the low-density neutron gas, as seen in Fig. \ref{fig:comparison}, we will 
also make calculations where we use the Skyrme EDF only for $P^{(l)}$ and $\mu_q^{(l)}$ in the liquid phase, while we use the BMBPT3 results for 
dilute neutron matter \cite{Palaniappan2023} to compute $P^{(g)}$ and $\mu_n^{(g)}$ in the gas phase. In practice, in order to compute $P^{(g)}$
and $\mu_n^{(g)}$ in this case, we approximate the numerical results of \cite{Palaniappan2023} by a rational function given in the appendix.

It is clear that this somewhat inconsistent approach only makes sense if the gas density is much smaller than the density in the liquid phase, and 
that it must fail if both phases have comparable densities, as it happens near the crust-core transition.

%%%%%%%%%%%%%%%%%%%%%%%%%%%%%%%%%%%%%%%%%%%%%%%%%%%%%%%%%%%%%%%%%%%%%%%%%%%%%%%%%%%%%%%%%%%%%%%%%%%%%%%%%%%%%%%%%%%%%%%%%%%%
\section{Results}
%%%%%%%%%%%%%%%%%%%%%%%%%%%%%%%%%%%%%%%%%%%%%%%%%%%%%%%%%%%%%%%%%%%%%%%%%%%%%%%%%%%%%%%%%%%%%%%%%%%%%%%%%%%%%%%%%%%%%%%%%%%%

\begin{figure*}
    \centering
    \includegraphics[width=\linewidth]{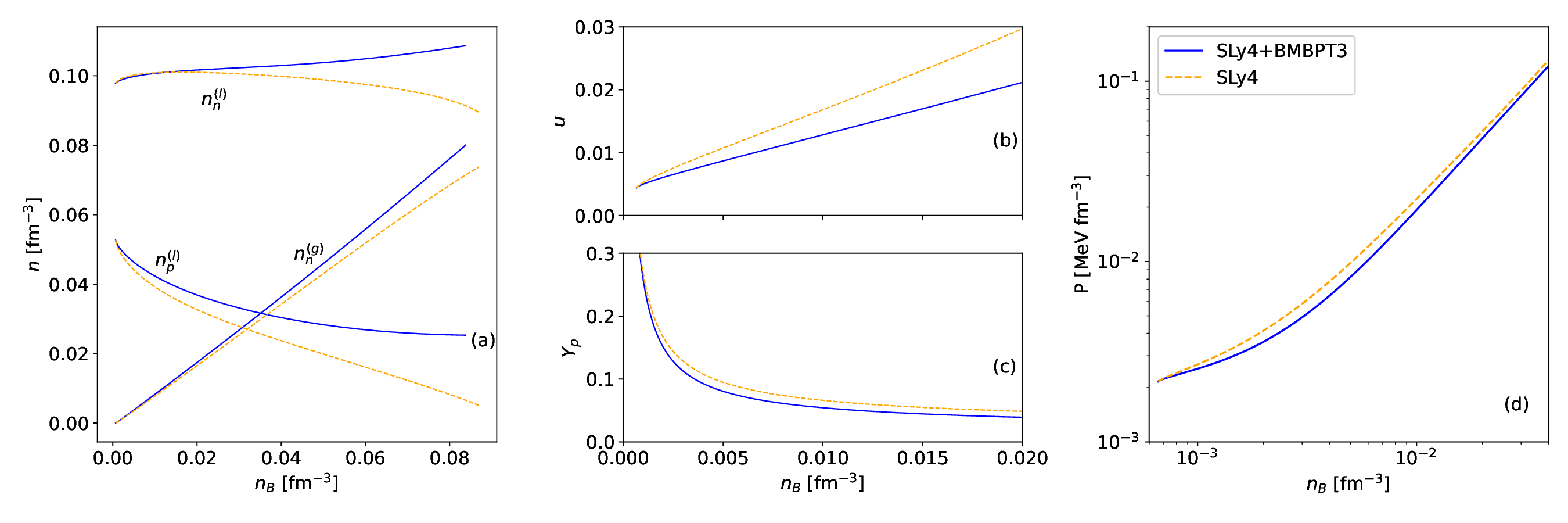}
\caption{(a) Comparison of different neutron and proton densities in the liquid and in the gas as functions of the total
baryon density, obtained after taking Skyrme interaction (SLy4) for both the clusters and the gas phase (orange lines)
and with taking Skyrme interaction (SLy4) for the clusters and realistic $nn$ interaction,(BMBPT3) for the gas phase
(blue lines). 
(b) Corresponding volume fractions of nuclear clusters as functions of total baryon density, 
(c) proton fraction,
(d) corresponding pressures, including the pressures of nucleons and electrons.}
\label{fig:graphab_sly}\end{figure*}

As explained at the end of the previous section, we compare results for the inner crust in the phase
coexistence model obtained in two different kinds of calculations to answer our question how employing realistic 
$nn$ interaction for neutron gas would change the composition, i.e., the densities $n_q^{(l)}$ and 
$n_n^{(g)}$ of the coexisting phases and the volume fraction $u$ of the liquid phase, i.e., of the nuclear clusters.First, we use the SLy4 interaction for the bound nuclei (liquid phase) and the BMBPT3 results for the neutron 
gas and then calculate the composition. This is compared with the results found with employing only the SLy4 
interaction for both phases. We can see the graph comparing the compositions obtained in these two 
calculations in 
Fig. \ref{fig:graphab_sly}(a).

%%%%%%%%%%%%%%%%%%%%%%%%%%%%%%%%%%%%%%%%%%%%%%%%%%%%%%%%%%%%%%%%%%%%%%%%%%%%%%%%%%%%%%%%%%%%%%%%%%%%%%%%%%%%%%%%%%%%%%%%%
%%%%%%%%%%%%%%%%%%%%%%%%%%%%%%%%%%%%%%%%%%%%%%%%%%%%%%%%%%%%%%%%%%%%%%%%%%%%%%%%%%%%%%%%%%%%%%%%%%%%%%%%%%%%%%%%%%%%%%%%%
A graph comparing 
the corresponding volume fractions of 
nuclear clusters as a function of $n_B$ is shown in Fig. \ref{fig:graphab_sly}(b), focusing on the low-density regime.

Looking at the curve of composition Fig. \ref{fig:graphab_sly}(a), we expect that the density of neutron gas goes to zero at
very low baryon density. This is the transition from outer crust to inner crust, which happens at the neutron-drip density $n_B^{\text{(ND)}}$. Our value $n_{B}^{\text{(ND)}} \approx 6.6\cdot 10^{-4}$ fm$^{-3}$ is approximately three times higher than the expected one \cite{Chamel2008}. The reason is that in our simple model we are neglecting the nuclear surface tension and Coulomb interaction. These can be taken into account in the framework of the compressible liquid-drop model that was introduced in \cite{Baym1971} and has since then been used by many authors, e.g., \cite{Vinas2017,Carreau2019,Dinh2021,DinhThi2023}. However, this goes beyond the scope of the present study.

Later on, as the total baryon density progressively increases as we go towards the core, the density of the neutron gas increases and nuclear 
clusters get enriched in neutrons, therefore the proton fraction decreases, see Fig. \ref{fig:graphab_sly}(c). When we take only 
SLy4 interaction for both phases, at $n_B \approx 0.09$ fm$^{-3}$ \cite{Martin2015}, the liquid fills the whole volume which is the 
transition from the inner crust to the outer core. Taking different interactions for liquid phase and gas phase we cannot describe this transition because of the inapplicability of this method at higher densities, as mentioned before.

In the low-density regime, when we employ BMBPT3 results to the gas phase, neutron and proton densities inside the nuclear clusters and the density of the neutron gas get larger, see Fig. \ref{fig:graphab_sly} (a). At the same time, since the total baryon density must be the same, the volume fraction of the clusters is reduced, see Fig. \ref{fig:graphab_sly} (b). Therefore, in total, we get 
more superfluid neutron gas content than when we apply only SLy4 interaction for both phases.
But since the volume fraction of the clusters is smaller with BMBPT3 than with SLy4 for the neutron gas, the net effect is that the proton fraction is reduced, see Fig. \ref{fig:graphab_sly} (c). Certainly we can say that our calculation with realistic $nn$ interaction gives us a hint that the inner crust is more neutron rich than predicted with the SLy4 interaction.
Now, making use of the compositions discussed above, we calculate the pressure of the system, including also the pressure of the degenerate electrons in addition to the pressure of the nucleons. In Fig. \ref{fig:graphab_sly} (d), we see that employing BMBPT3 in the gas phase reduces the pressure of the system as compared to only SLy4 interaction in both phases.

\begin{figure*}
    \includegraphics[width = \linewidth]{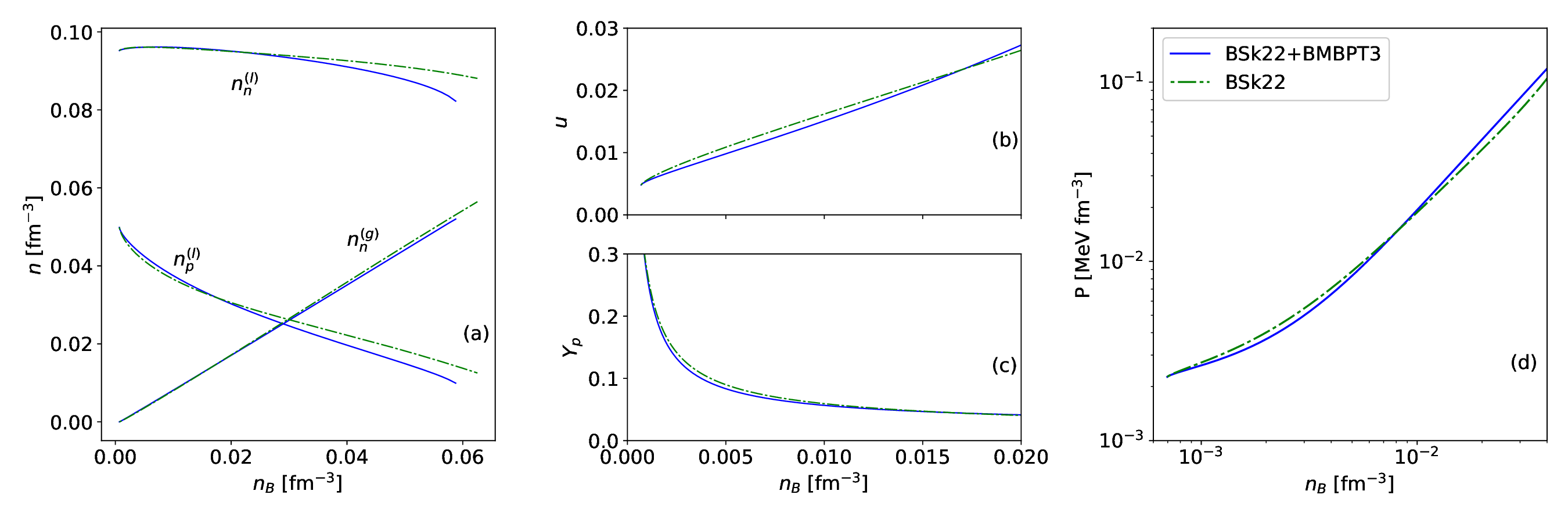}
\caption{Same as Fig. \ref{fig:graphab_sly} but for BSk22 interaction instead of SLy4.}
\label{fig:graphab_bsk}\end{figure*}

%explaining fig. (3) BSk22 and BSk22+BMBPT3
In the next step we will incorporate another phenomenological model BSk22 in place of SLy4 and repeat our study. We take BMBPT3 for the gas phase and BSk22 for the nuclear clusters and compare with the only BSk22 interaction in both phases. We notice that now the curves
of the composition are similar in both cases, see Fig. \ref{fig:graphab_bsk} (a). Eventually the volume fraction, proton fraction, and pressure, see Figs. \ref{fig:graphab_bsk} (b), (c), and (d), turn also out to be very similar in both cases.
%%%%%%%%%%%%%%%%%%%%%%%%%%%%%%%%%%%%%%%%%%%%%%%%%%%%%%%%%%%%%%%%%%%%%%%%%%%%%%%%%%%%%%%%%%%%%%%%%%%%%%%%%%%%%%%%%%%%%%%%%
%%%%%%%%%%%%%%%%%%%%%%%%%%%%%%%%%%%%%%%%%%%%%%%%%%%%%%%%%%%%%%%%%%%%%%%%%%%%%%%%%%%%%%%%%%%%%%%%%%%%%%%%%%%%%%%%%%%%%%%%%
\section{Conclusion}
Looking at Fig. \ref{fig:comparison}, we ascertain that phenomenological Skyrme EDFs disagree with the ab-initio (BMBPT3) EoS based on a realistic $nn$ interaction in the low-density regime of neutron matter, as it can be found in the inner crust of neutron stars. Therefore we study how strongly this discrepancy may affect predictions for the composition of the inner crust based on Skyrme interactions. To get an idea of the importance of the effect, we consider the simple phase-coexistance model and compare the results obtained by using the phenomenological Skyrme EDF for the nuclear clusters and the BMBPT3 EoS for the neutron gas with those obtained by using the Skyrme EDF for both phases.

We find out that using only SLy4, for gas phase and nuclear clusters, is not in agreement with the calculation done with the BMBPT3 employed for gas phase and SLy4 for nuclear clusters.
Replacing SLy4 with BSk22 and repeating our calculation, we observe that the composition of the inner crust found using only BSk22 for both phases turns out to be similar to that obtained with BMBPT3 for the gas phase and BSk22 for the nuclear clusters. Referring to Fig. \ref{fig:comparison}, we could have anticipated that the discrepancy in the BSk22 case is smaller than in the SLy4 case, since the behavior of BSk22 is overall closer to BMBPT3 than SLy4 and almost coincides with BMBPT3 at $k_F \gtrsim 0.8$ fm$^{-1}$.  

We conclude that, in order to obtain a reasonable description of the inner crust, it is sufficient to employ an EDF such as BSk22 that follows closely the correct neutron-matter EoS at $n_n\gtrsim 0.02$ fm$^{-3}$, even if it does not have the right asymptotic behavior for $n_n\to 0$. This is consistent with the result of Ref. \cite{Shelley2020} where it was found that the inclusion of pairing, which affects the neutron matter EoS in the same density range where BSk22 disagrees with BMBPT3, has almost no effect on the inner crust composition.

In this work, we implemented the simple phase coexistence model in order to study the composition of the inner crust and the behaviour of various phenomenological Skyrme EDFs and ab-initio theory at low density. While this is probably sufficient to estimate the effect of the differences between the various EoS, it is clearly not enough to make a reliable prediction for the crust composition. For instance, the neutron drip density is not correct. Therefore,
in further studies, we wish to extend our calculation to the compressible liquid-drop model, where we can include surface and Coulomb energy in our calculations, getting closer to the real physical system. Furthermore, in the Skyrme EoS, the effect of $nn$ pairing should be included.

\appendix
\section{Rational approximation for the neutron-gas EoS}
\label{rational_approximation_BMBPT3}

We made a rational approximation of the BMBPT3 results of Ref. \cite{Palaniappan2023} which is useful to compute the chemical potential and pressure of the dilute neutron gas.
We write the ratio as
\begin{equation}
    \frac{\eps}{\eps_{\text{FG}}} (\kF) = \frac{1 +a_1 \kF+a_2 \kF^2+a_3 \kF^3+a_4 \kF^4}{1+b_1 \kF+b_2 \kF^2 + b_3 \kF^3+ b_4 \kF^4}
    \label{rational_eqn} \,.
\end{equation}
When the density tends towards zero, this ratio should behave as \cite{Fetter-Walecka}
\begin{equation}
    \frac{\eps}{\eps_{\text{FG}}} = 1 + \frac{10}{9 \pi} \kF a_{nn} \,,
    \label{ratioatzero}
\end{equation}
where $a_{nn} = -18.487$ fm is the $nn$ scattering length (see Table VIII of \cite{Wiringa1995}).
From Eq. (\ref{ratioatzero}), one can obtain $b_1 = a_1-\frac{10}{9 \pi}a_{nn}$. The
other parameters are obtained by a least-square fit and are listed in Table \ref{tab:parameters_rational}. 
The parameters $a_2$ and $b_3$ have been set to zero because they do not lead to any improvement of the fit. 
%%%%%%%%%%%%%%%%%%%%%%%%%%%%%%%%%%%%%%%%%%%%%%%%%%%%%%%%%%%%%%%%%%%%%%%%%%%%%%%%%%%%%%%%%%%%%%%%%%%%%%%%%%%%%%%%%%%%%%%%
\begin{table}[t]
    \begin{ruledtabular}
    \begin{tabular}{cccc}
        $a_1$ (fm) & $a_2$ (fm$^2$) & $a_3$ (fm$^3$) & $a_4$ (fm$^4$) \\
        7.42207    & 0              &16.0561         &2.09893 \\
        \hline
        $b_1$ (fm)                    & $b_2$ (fm$^2$) & $b_3$ (fm$^3$) & $b_4$ (fm$^4$) \\
        13.9605    &9.55301         & 0              &35.3716
    \end{tabular}
    \end{ruledtabular}
    \caption{Parameters of the rational approximation (\ref{rational_eqn}).}
    \label{tab:parameters_rational}
\end{table}
%%%%%%%%%%%%%%%%%%%%%%%%%%%%%%%%%%%%%%%%%%%%%%%%%%%%%%%%%%%%%%%%%%%%%%%%%%%%%%%%%%%%%%%%%%%%%%%%%%%%%%%%%%%%%%%%%%%%%%%%

\section{Pairing energy}\label{app:pairing}

To estimate the pairing energy to be added to the phenomenological Skyrme EoS, it is important to include screening (or medium polarization) effects which are known to lower the pairing gap
compared to when using only the bare $nn$ interaction in the pairing channel. Here, we use the pairing interaction as described in Ref.~\cite{Urban2020},
including the full RPA polarization computed with SLy4 and BSk22, respectively. Gaps computed with this interaction have the correct asymptotic behaviour
at very low density and also agree well with Quantum Monte-Carlo results, cf. \cite{Ramanan2020}. In BSk22, screening turns out to be weaker so that the gaps 
are somewhat larger than with SLy4. The pairing energy density is computed from the formula \cite{Fetter-Walecka}
\begin{equation}
\varepsilon_{\text{pair}} = \int \frac{d^3k}{(2\pi)^3} \Big(\frac{\hbar^2k^2}{m^*}[v_k^2-\theta(k_F-k)]-u_k v_k\Delta_k\Big),
\label{eq:pairingenergy}
\end{equation}
with $m^*$ the effective mass computed with the respective Skyrme functional, and $u_k$ and $v_k$ the usual coefficients of the Bogoliubov transformation and $\Delta_k$ the momentum dependent gap, cf. \cite{Fetter-Walecka}.

\bibliography{bibliography}

\end{document}